\documentclass[cameraready]{Interspeech}

\title{Diffusion Reconstruction towards Generalizable Audio Deepfake Detection}

\author[affiliation={1, *}]{Bo}{Cheng}
\author[affiliation={2}]{Songjun}{Cao}
\author[affiliation={2}]{Xiaoming}{Zhang}
\author[affiliation={2}]{Jie}{Chen}
\author[affiliation={2}]{Long}{Ma}
\author[affiliation={1, **}]{Fei}{Chen}

\address{
    $^1$ Department of Electronic and Electrical Engineering, \\
Southern University of Science and Technology, Shenzhen, China \\
    $^2$ Tencent Youtu Lab, China
}

\email{12442015@mail.sustech.edu.cn, fchen@sustech.edu.cn}

\keywords{audio deepfake detection, generalization, audio reconstruction}

\usepackage{comment}

\begin{document}

\maketitle

\begin{abstract}
Achieving robust generalization against unseen attacks remains a challenge in Audio Deepfake Detection (ADD), driven by the rapid evolution of generative models. 
To address this, we propose a framework centered on hard sample classification. 
The core idea is that a model capable of distinguishing challenging hard samples is inherently equipped to handle simpler cases effectively. 
We investigate multiple reconstruction paradigms, identifying the diffusion-based method as optimal for generating hard samples. 
Furthermore, we leverage multi-layer feature aggregation and introduce a Regularization-Assisted Contrastive Learning (RACL) objective to enhance generalizability. 
Experiments demonstrate the superior generalization of our approach, with our best model achieving a significant reduction in the average Equal Error Rate (EER) compared to the baseline. 
\end{abstract}

\section{Introduction}
In recent years, the rapid progress of deep learning techniques has made AI-generated content more realistic. 
Driven by advanced architectures, contemporary Text-to-Speech (TTS) \cite{zhang25c_interspeech, wu25g_interspeech} and Voice Conversion (VC) \cite{10888535, su25c_interspeech} make it possible to generate high-quality speech that is virtually indistinguishable to the human.
While the technology is a powerful tool for rapid creation and improving accessibility,
it presents significant threats to modern society if misused or not well-controlled.
Such technology may lead to severe security risks, such as telecommunications fraud and disinformation spreading.
Consequently, developing highly reliable ADD systems is imperative to ensure security.

\newcommand\blankfootnote[1]{%
  \begingroup
  \renewcommand\thefootnote{}\footnote{#1}%
  \addtocounter{footnote}{-1}%
  \endgroup
}
\blankfootnote{$^\ast$Work is done during the internship at Tencent Youtu Lab}
\blankfootnote{$^\ast$$^\ast$Corresponding author}

To detect deepfake audio, researchers have explored various detection strategies.
For instance, \cite{10889022} emphasizes the role of spectral representations, proposing methodologies to capture subtle discrepancies between bona fide and spoof signals within the spectrogram domain for robust classification.
Alternatively, \cite{10889337} targets global and local artifacts via spectro-temporal cross-aggregation and dynamic convolution, leveraging cross-attention to fuse temporal and frequency representations for effective synthetic speech discrimination.
Furthermore, \cite{10888328} enhances generalization against unseen attacks by integrating Latent Space Refinement and Augmentation, utilizing learnable prototypes and latent-level manipulations to diversify feature representations for detection.
Although these methods yield powerful ADD models, their limited capacity to generalize to unseen or cross-domain attacks remains a persistent challenge.
Given the continuous emergence of novel attacks, existing ADD approaches frequently struggle with out-of-domain audio, making it critical to improve model generalization.

Inspired by generated image detection \cite{pmlr-v235-chen24ay}, we address the challenge by focusing on hard sample classification. 
As discussed in \cite{pmlr-v235-chen24ay}, a model that can discriminate between hard samples is, by definition, competent at classifying trivial ones.
Our framework leverages the reconstruction mechanism to generate hard samples. 
Similar to observations in the image domain, reconstructed audio sounds identical to the source recording. 
However, it contains the subtle artifacts inherent to generative models.
We integrate multi-layer feature aggregation with RACL to enhance the model's generalization capability against unseen attacks.
Experimental analysis over various reconstruction paradigms shows that diffusion-based reconstruction offers the strongest generalization across diverse attack types.
The contributions of this paper are summarized as follows:

\begin{itemize}
\item \textbf{Hard Sample Construction via Reconstruction:}
Following a comprehensive evaluation of various reconstruction methods, we determine that the diffusion-based approach yields the most effective results.
Our proposed method achieves an average EER of $12.220$\% across five test sets, representing a relative reduction of $22.604$\% compared to the baseline's average of $15.789$\%.

\item \textbf{Regularization-Assisted Contrastive Learning:} 
We design a novel optimization objective that combines a dual contrastive loss (including standard \cite{1640964} and enhanced versions) with a regularization loss. 
Through ablation studies, we validate that both components are effective in enhancing the model's generalization capability.

\end{itemize}

\begin{figure*}[t]
  \centering
  \includegraphics[width=\textwidth]{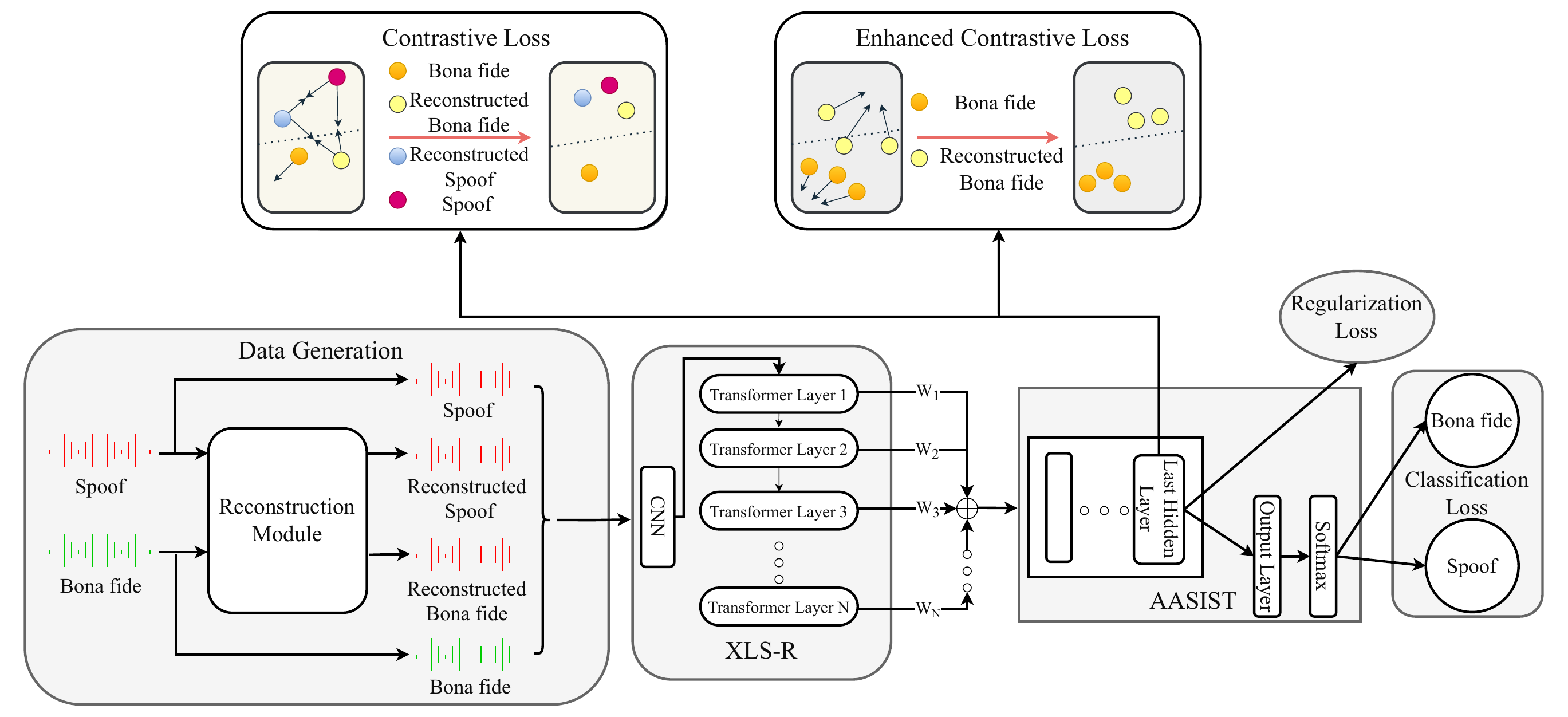}
  \caption{
  The workflow of the proposed framework.
  It integrates audio reconstruction module with an ADD module comprising a pretrained XLS-R 300M and an AASIST. 
  During the training phase, the model learns from all bona fide, spoof, and reconstructed samples. 
  We optimize the network by using the RACL.
  }
  \label{fig:workflow}
\end{figure*}

\section{Method}

\subsection{Overview}
As illustrated in Figure \ref{fig:workflow}, we first reconstruct bona fide and spoof audio using diverse models (HiFi-GAN \footnote{\url{https://github.com/jik876/hifi-gan?tab=readme-ov-file}} \cite{10.5555/3495724.3497152}, DAC \footnote{\url{https://github.com/descriptinc/descript-audio-codec}}  \cite{10.5555/3666122.3667336},  Encodec \footnote{\url{https://github.com/facebookresearch/encodec}} \cite{defossez2022highfi} and SemantiCodec \cite{semanticodec2024}) to generate hard samples.
For detection, features are extracted by a frozen XLS-R 300M \footnote{\url{https://github.com/facebookresearch/fairseq}}  \cite{babu22_interspeech} and then processed by an AASIST \cite{9747766} for classification.
In addition, we apply the \textbf{RACL} to the embeddings from the last hidden layer of AASIST to enhance generalization capability.
         
\subsection{Data Generation}
We employ several models to generate the reconstructed samples.
For HiFi-GAN, we first extract mel-spectrograms from the original audio and subsequently resynthesize the waveforms using the pretrained vocoder.
Conversely, for Encodec and DAC, we directly process raw audio using the pretrained models to generate reconstructed waveforms.
In particular, SemantiCodec \cite{semanticodec2024} demonstrates superior reconstruction performance among these methods.
Briefly, it extracts semantic and acoustic features via a dual-encoder architecture, which serve as conditions for a Latent Diffusion Model (LDM) to predict latent representations.
To generate the final waveform, these predicted latents are passed through a decoder and subsequently processed by a vocoder.
We utilize the official pretrained checkpoint \footnote{\url{https://huggingface.co/haoheliu/SemantiCodec/tree/main}} for diffusion-based implementation.

\subsection{Model Architecture}
Our framework integrates a pretrained XLS-R 300M with an AASIST.
We implement an adaptive layer aggregation module \cite{Wang_2020_CVPR, 10.1145/3664647.3681345} to compute a weighted summation of the outputs from all $L$ transformer layers.
Let $\mathbf{F}_l \in \mathbb{R}^{T \times D}$ denote the output of the $l$-th layer of XLS-R. 
We first compress each layer into a scalar descriptor $z_l$ via Global Average Pooling (GAP):
\begin{align}
    z_l = \frac{1}{T \times D} \sum_{i=1}^{T} \sum_{j=1}^{D} \mathbf{F}_l(i, j).
  \label{equation:eq4}
\end{align}
The resulting vector $\mathbf{z} = [z_1, \dots, z_L]$ is processed via a 1D convolution with adaptive kernel size $k$ to yield attention weights $\boldsymbol{\omega}$:
\begin{align}
    \boldsymbol{\omega} = \sigma(\text{Conv1D}_k(\mathbf{z})).
    \label{equation:eq_weights}
\end{align}
Finally, the aggregated feature $\mathbf{F}_{agg} = \sum_{l=1}^{L} \omega_l \cdot \mathbf{F}_l$ is fed into the AASIST for classification.

\subsection{Loss Function}

\subsubsection{Dual Contrastive Loss}
To strictly distinguish bona fide samples, we propose a Dual Contrastive Loss $\mathcal{L}_{dual} = \alpha \mathcal{L}_{std} + \beta \mathcal{L}_{enh}$. 
Both losses adopt the margin-based contrastive formulation \cite{1640964}
\begin{align}
    \begin{gathered}
        \mathcal{L} = \frac{1}{N} \sum_{i=1}^{N} \Big[ Y \cdot D_w(i)^2 \\
        + (1 - Y) \cdot \max(0, m - D_w(i))^2 \Big]
    \end{gathered}
\end{align}
where $D_w(i)$ denotes the Euclidean distance and $Y$ represents the binary pair label. 
Specifically, we define $Y = 1$ if the two samples share the same classification label (i.e., both are bona fide or both are spoof), and $Y = 0$ otherwise.  
Following \cite{pmlr-v235-chen24ay}, we employ $\mathcal{L}_{std}$ as our standard contrastive loss. 
However, $\mathcal{L}_{std}$ assigns equal weighting to all samples, lacking the sensitivity required to discern hard sample.
We therefore propose $\mathcal{L}_{enh}$ to emphasize the discrimination of the hard sample.
Unlike $\mathcal{L}_{std}$ which operates on all samples, $\mathcal{L}_{enh}$ focuses exclusively on bona fide and reconstructed bona fide samples.

\begin{table*}[t]
  \caption{
    The best performance is highlighted in bold. 
    Notations: 
    \textbf{Baseline$^*$} refers to CodecFake \cite{10830534}. 
    \textbf{Baseline} is our implementation. 
     \textbf{HiFi-GAN}, \textbf{DAC}, \textbf{Encodec} and \textbf{Diffusion} utilize HiFi-GAN, DAC,  Encodec and SemantiCodec for data reconstruction, respectively. 
    \textbf{Agg Diffusion} utilizes multi-layer aggregation with diffusion reconstruction.
    \textbf{RACL Diffusion} integrates multi-layer aggregation, RACL, and diffusion reconstruction. }
  \label{tab:all}
  \centering
  \begin{tabular}{l c c c c c c c c}
    \toprule
    \textbf{Test Set} & \textbf{Baseline$^*$} & \textbf{Baseline} & \textbf{HiFi-GAN} & \textbf{DAC} & \textbf{Encodec} & \textbf{Diffusion} & \textbf{Agg Diffusion} & \textbf{RACL Diffusion} \\
    \midrule
    ASVspoof          & 0.122  & 0.216  & 0.201  & 1.010  & 0.295  & \textbf{0.166}  & 0.288  & 0.206 \\
    ITW               & 23.713 & 17.949 & 23.779 & 39.477 & 22.964 & 18.159                  & 10.679 & \textbf{9.155} \\
    DiffSSD           & --     & 21.587 & 38.991 & 25.833 & 15.129 & 14.479                  & 10.446 &  \textbf{10.081} \\
    WaveFake          & --     & 2.395  & 1.723  & 3.319  & 3.031  & \textbf{1.235}          & 1.968  & 1.597 \\
    CodecFake         & 41.583 & 36.799 & 39.616 & 39.972 & 29.816 & 27.063                  & 21.061 & \textbf{20.198} \\
    \midrule
    \textbf{Avg Total} & --    & 15.789 & 20.862 & 21.922 & 14.247 & 12.220                  & 8.888  & \textbf{8.247} \\
    \bottomrule
  \end{tabular}
\end{table*}

\subsubsection{Regularization Loss}
Inspired by the regularization strategy proposed in \cite{10096915}, we employ a variance-based loss to enforce intra-class compactness for both bona fide and spoof classes.
In contrast to the original objective, we aim to enforce tighter, more cohesive clustering at the batch level. The regularization loss is therefore defined as follows:

\begin{align}
    \mathcal{L}_{reg} = - \frac{1}{d} \sum_{j=1}^{d} \max \left( 0, 1 - \sqrt{\text{Var}(\mathbf{x}_{j}) + \delta} \right)
\end{align}
Here, $d$ denotes the feature dimension, $\text{Var}(\mathbf{x}_{j})$ denotes the variance of the $j$-th dimension, and $\delta$ is a negligible constant. 
This loss is computed independently for the bona fide and non-bona fide classes, with the latter encompassing all three spoof categories, to minimize the feature-wise variance of embeddings within each batch. 
By constraining the variance along each dimension toward zero, the objective encourages the distributions to aggregate densely, thereby yielding more compact and cohesive intra-class clusters.


\subsubsection{Classification Loss}
We adopt the cross-entropy loss to distinguish between bona fide and all other sample types.
The loss function is defined as:
\begin{align}
    \mathcal{L}_{cls} = - \sum_{i=1}^{M} \Big[ y_i \log(p_i) + (1 - y_i) \log(1 - p_i) \Big]
\end{align}
where $p_i$ denotes the predicted probability of the sample belonging to the non-bona fide class, and $y_i$ is the ground truth label.
Specifically, we define the label $y_i = 0$ exclusively for bona fide samples, while assigning $y_i = 1$ to all other variations, including reconstructed bona fide samples, spoof samples, and reconstructed spoof samples.

\subsubsection{Regularization-Assisted Contrastive Learning}
We formulate the RACL objective as a weighted combination of classification, contrastive, and regularization losses:
\begin{align}
    \mathcal{L}_{total} = (1 - \alpha - \beta) \mathcal{L}_{cls} + \mathcal{L}_{dual} + \gamma \mathcal{L}_{reg}
\end{align}
where $\mathcal{L}_{dual} = \alpha \mathcal{L}_{std} + \beta \mathcal{L}_{enh}$. 
The coefficients $\alpha$, $\beta$, and $\gamma$ are hyperparameters, which are set to 0.6, 0.1, and 0.3.

\section{Experiments}
\subsection{Dataset}
We evaluate our models on five diverse and comprehensive datasets. 
\textbf{ASVspoof 2019 LA eval} \cite{Nautsch2021ASVspoof2S} comprises TTS and VC utterances synthesized via traditional vocoders, whereas \textbf{CodecFake} \cite{10830534} focuses on deepfakes derived from neural audio codecs.
Crucially, to prevent data leakage, the CodecFake test set excludes bona fide samples that overlap with the training and validation partitions of \textbf{ASVspoof 2019 LA}. 
\textbf{DiffSSD} \cite{10889450} evaluates generalization against diffusion-based synthesis using 8 diffusion methods and 2 commercial APIs.
\textbf{WaveFake} \cite{frank2021wavefake} assesses the detection performance of GAN-based synthesis.
\textbf{ITW} \cite{muller22_interspeech} collected audio from social media to test performance in uncontrolled environments.

\begin{table}[htbp]
  \caption{
  Detailed EER (\%) comparison on CodecFake subsets.
  }
  \label{tab:codecfake_subsets}
  \centering
  \begin{tabular}{cccccc}
    \toprule
    \textbf{Subset} & \textbf{Baseline} & \textbf{DAC} & \textbf{Encodec} & \textbf{Diffusion} \\
    \midrule
    C1   & 29.009 & 69.081 & 25.111 & \textbf{19.956} \\
    C2   & 50.791 & 72.004 &54.799 & \textbf{37.885} \\
    C3   & 37.782 & 26.841 & 25.598 & \textbf{24.624} \\
    C4   & 25.393 & 24.482 & \textbf{7.221} & 16.244 \\
    C5   & 27.014 & 25.193 & 17.334 & \textbf{15.183} \\
    C6   & 44.238 & \textbf{33.813} & 40.411 & 38.471 \\
    C7   & 43.367 & \textbf{28.387} & 38.239 & 37.077 \\
    \midrule
    \textbf{Average} & 36.799 & 39.972 & 29.816 & \textbf{27.063} \\
    \bottomrule
    \end{tabular}
\end{table}

\subsection{Implementation Details}
All audio samples across the training, development, and evaluation sets are resampled to \SI{16}{\kilo\hertz} and normalized to a fixed duration of 64,600 samples. 
Waveforms exceeding this length are truncated, while shorter ones are extended via circular padding. 
For data augmentation, we utilize RIRs \cite{7953152} and the MUSAN corpus \cite{musan2015}. 
Specifically, noise and music components are added with Signal-to-Noise Ratios (SNR) uniformly sampled from intervals of $[0, 15]$ dB and $[5, 15]$ dB, respectively. 
For speech augmentation, we mix 3 to 8 randomly selected utterances, with SNR sampled from $[13, 20]$ dB.

The proposed system integrates a frozen XLS-R with a trainable AASIST.
Training was conducted for 100 epochs with a fixed random seed of 688.
We optimized the AASIST using Adam ($\beta_1=0.9, \beta_2=0.999, \epsilon=1 \times 10^{-8}$) with a weight decay of $5 \times 10^{-4}$. 
The cross-entropy loss was configured with class weights of $[10, 1]$ for the bona fide and spoof classes. 
The initial learning rate is set to $5 \times 10^{-4}$ and decays by a factor of 0.5 every 10 epochs. 
The final model is obtained by averaging the parameters of the checkpoint with the lowest validation loss and its four preceding epochs.

\subsection{Results}
\subsubsection{Improving Generalization via Reconstruction}
We employed five distinct methods (detailed in Table \ref{tab:all}) to reconstruct both bona fide and spoof utterances from the ASVspoof 2019 LA training and development sets, while the baseline utilized only the original ASVspoof 2019 LA data. 
To ensure the reliability of our results, all experiments were independently repeated three times, with the mean performance results reported.

Table \ref{tab:all} presents the detailed performance comparison. 
Results indicate that:
(1) The inclusion of reconstructed data does not inherently ensure performance enhancements across every individual test set.
(2) Albeit effective for datasets produced by analogous generative models, particular reconstruction paradigms often lack the capacity to generalize to unseen spoofing attacks.
(3) The diffusion-based reconstruction achieves consistent performance gains across all test sets and outperforms other reconstruction approaches. 
We attribute the results to the stochastic nature of diffusion-based synthesis, which effectively simulates complex real-world scenarios, thereby boosting global generalization.
(4) The proposed framework, integrating multi-layer aggregation with the proposed RACL, achieves a significant reduction in the average EER results.
Minor performance degradation occurs in some datasets.
This suggests that the model prioritizes generalized features over specific artifacts, achieving a better global optimum.

Tables \ref{tab:codecfake_subsets} provides a comprehensive and detailed breakdown of performance across specific attack subsets.
Within the CodecFake dataset, the Encodec-based approach yields the most favorable detection results on C4 (Encodec-generated), while the DAC-based method excels particularly on the DAC-generated C7.
However, the diffusion-based reconstruction paradigm secures the highest overall average performance and consistently exceeds the vanilla baseline across all subsets. 
Collectively, these results highlight the superior and robust generalization capabilities of the diffusion-based approach, empirically validating our earlier conclusions (1), (2), and (3).



\subsubsection{Ablation Experiments}
Table \ref{tab:ablation_loss} presents the ablation study on the RACL.
The results demonstrate that incorporating $\mathcal{L}_{enh}$ yields superior performance compared to the configuration without it. 
This improvement stems primarily from $\mathcal{L}_{enh}$, which effectively enforces the separation of hard samples, thereby enhancing the model's generalization capability.
The integration of $\mathcal{L}_{reg}$ fosters performance enhancements by minimizing intra-class feature distances.
Optimal results are attained when all loss components are utilized in concert. 
Within this configuration, $\mathcal{L}_{reg}$ serves as a \textit{stabilizer}.
By anchoring intra-class variance, it enables $\mathcal{L}_{enh}$ to refine decision boundaries without compromising the underlying feature structure.

\begin{table}[htbp]
\centering
\caption{
Ablation study on ${\mathcal{L}_{std}}$, ${\mathcal{L}_{enh}}$, and ${\mathcal{L}_{reg}}$.
The results are reported in terms of average EER (\%) of five datasets. 
The best performance is highlighted in bold.}
\label{tab:ablation_loss}
\begin{tabular}{ccccc}
\toprule
\textbf{${\mathcal{L}_{cls}}$} & \textbf{${\mathcal{L}_{std}}$} & \textbf{${\mathcal{L}_{enh}}$} & \textbf{${\mathcal{L}_{reg}}$} & \textbf{Avg EER} \\ 
\midrule
\checkmark & & & & 10.328 \\ 
\checkmark & \checkmark & & & 8.888 \\ 
\checkmark & \checkmark & \checkmark & & 8.640 \\
\checkmark & \checkmark & \checkmark & \checkmark & \textbf{8.247}\\
\bottomrule
\end{tabular}
\end{table}

\begin{figure}[t]
  \centering
  \includegraphics[width=\linewidth]{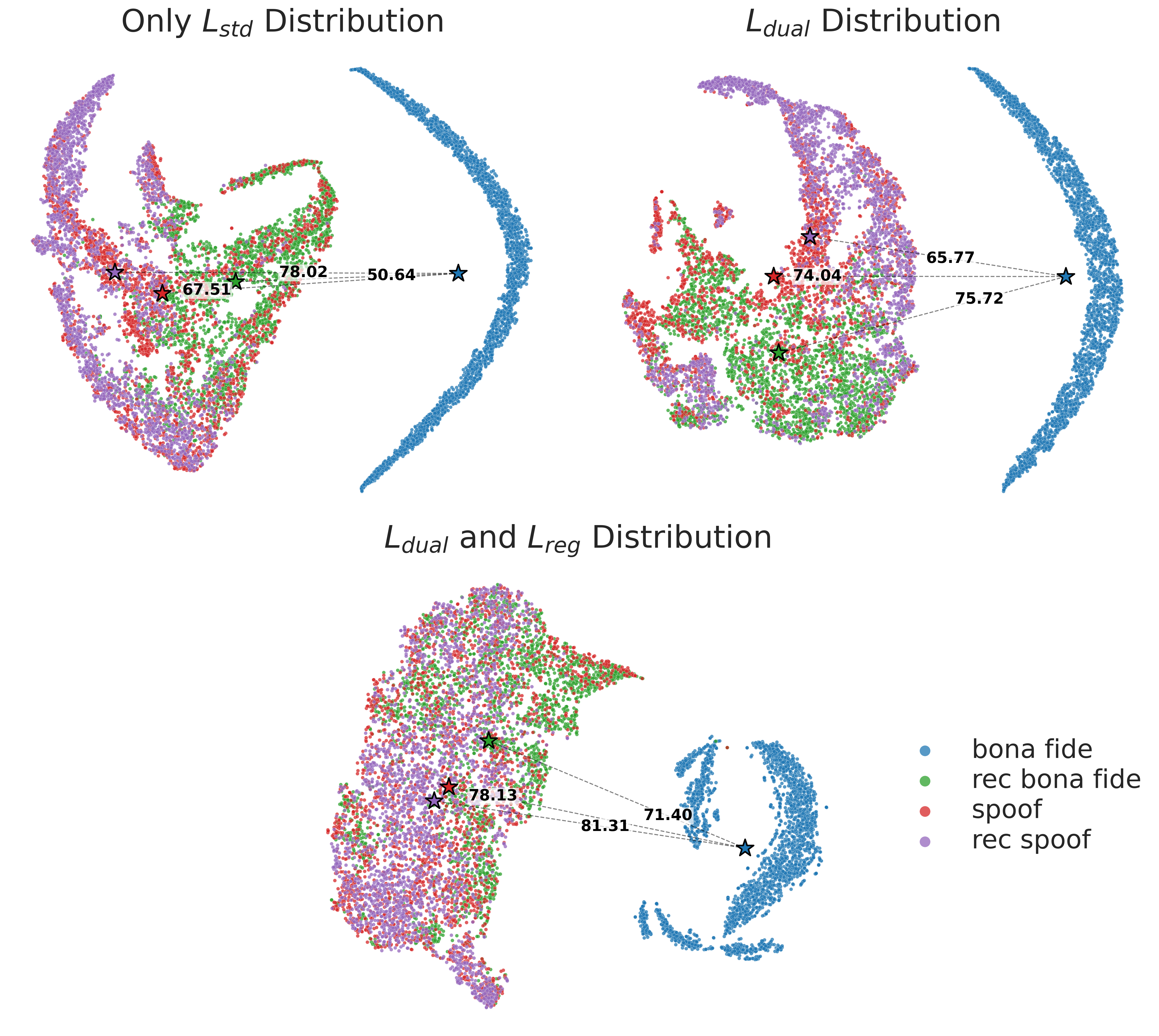}
    \caption{
    t-SNE visualization of feature embeddings. 
    The plots correspond to the configurations in the second (top-left), third (top-right), and final (bottom) rows of Table \ref{tab:ablation_loss}.}
    \label{fig:tsne}
\end{figure}

\subsubsection{Visualization}
Figure \ref{fig:tsne} presents the t-SNE visualization of embeddings extracted from the last hidden layer of the AASIST. 
Specifically, the visualization dataset consists of 2580 original bona fide utterances, 2580 randomly sampled spoof utterances, and their corresponding reconstructed counterparts (denoted as \textit{bona fide}, \textit{spoof}, \textit{rec bona fide}, and \textit{rec spoof} in Figure \ref{fig:tsne}).
Quantitative analysis reveals that $\mathcal{L}_{enh}$ increases the average distance between bona fide samples and other categories from 65.39 to 71.84. 
Notably, the separation between original and reconstructed bona fide samples expands to 75.72.
It is greater than the baseline distance of 50.64, which confirms that $\mathcal{L}_{enh}$ effectively forces the discrimination of hard samples. 
Furthermore, incorporating $\mathcal{L}_{reg}$ maximizes the overall distance to 76.95 and significantly compacts intra-class distributions. 
These findings demonstrate that the proposed RACL enhances feature compactness while widening the margin between bona fide audio and its reconstructed counterparts.

\section{Conclusion}
In summary, through a comprehensive comparison of various reconstruction paradigms, we demonstrate that the diffusion-based strategy yields significant performance improvements across diverse datasets. 
Crucially, the model achieves consistent gains over the baseline across all individual spoofing methods in the evaluated subsets. 
These results provide compelling evidence for the enhanced generalization capabilities of the diffusion-based reconstruction.
Finally, the ablation experiments and feature visualization confirm that the proposed RACL significantly strengthen the generalizability of our ADD model.

\section{Generative AI Use Disclosure}
We utilized generative AI tools to refine the linguistic presentation of this manuscript.

\bibliographystyle{IEEEtran}
\bibliography{mybib}

\end{document}